\documentclass{ws-ijmpd}

\usepackage{amsmath,amssymb}
\usepackage{graphicx}
\usepackage{epsfig}
\usepackage{float}
\usepackage{subfigure}
\usepackage{wrapfig}
\usepackage{multirow}
\usepackage{tabularx}
\usepackage{array}
\usepackage{url}
\usepackage{systeme}
\usepackage{braket}
\usepackage{xcolor}
\usepackage[super]{cite}
\usepackage[verbose,hypertexnames=false]{hyperref}
\hypersetup{
  colorlinks=false,
  allbordercolors=blue
}

\begin{document}

\markboth{Albino \textit{et al.}}{Sharpness of the quark-hadron transition and hybrid stars}

%
\catchline{}{}{}{}{}
%

\title{The sharpness of the quark-hadron transition and the properties of hybrid stars}

\author{M.~B.~Albino\footnote{milena.albino@student.uc.pt}}
\address{Departamento de Física Nuclear, Instituto de Física, Universidade de São Paulo,\\
Rua do Matão, 1371, CEP 05508-090, São Paulo, SP, Brazil.}

\author{R.~Fariello\footnote{fariello@if.usp.br}}
\address{Departamento de Física Nuclear, Instituto de Física, Universidade de São Paulo,\\
Rua do Matão, 1371, CEP 05508-090, São Paulo, SP, Brazil.}
\address{Departamento de Ciências da Computação, Universidade Estadual de Montes Claros,\\
Avenida Rui Braga, sn, Vila Mauricéia, CEP 39401-089, Montes Claros, MG, Brazil.}

\author{G.~Lugones\footnote{german.lugones@ufabc.edu.br}}
\address{Universidade Federal do ABC, Centro de Ci\^encias Naturais e Humanas,\\
Avenida dos Estados 5001\,–\,Bangu, CEP 09210-580, Santo André, SP, Brazil.}

\author{F.~S.~Navarra\footnote{navarra@if.usp.br}}
\address{Departamento de Física Nuclear, Instituto de Física, Universidade de São Paulo,\\
Rua do Matão, 1371, CEP 05508-090, São Paulo, SP, Brazil.}

\maketitle

\begin{abstract}
We investigate the effects of the sharpness of the phase transition between hadronic matter and quark matter on various properties of neutron stars. We construct hybrid equations of state by combining a hadronic model with a quark model using a Gaussian function. This approach introduces a smooth transition characterized by two parameters: one representing the overpressure relative to the first-order phase transition point, and the other related to the range over which the hybrid region extends in baryon chemical potential.
We find that the sharpness of the phase transition significantly influences the equation of state, which can deviate by several tens of $\text{MeV fm}^{-3}$ from the one with a sharp first-order transition. The speed of sound exhibits diverse behaviors—including drastic drops, pronounced peaks, and oscillatory patterns—depending on the sharpness parameters.
In terms of stellar structure, while the maximum neutron star mass remains largely unaffected by the sharpness of the phase transition, the stellar radii can vary significantly. Smoother transitions lead to a leftward shift (up to 1 km) of the mass-radius curve segment corresponding to hybrid stars. The tidal deformability decreases with smoother transitions, especially for higher-mass stars. Our results are quite general and do not qualitatively depend on the specific hadronic and quark matter models employed. In fact, the hybrid equation of state and stellar properties derived from microscopic models of quark-hadron pasta phases display the same behavior as described above. 
\end{abstract}

\maketitle

\section{Introduction}

The equation of state (EOS) of strongly interacting matter is relatively well known at low densities \cite{Oertel:2016bki,Tews:2012fj} and at ultrahigh densities \cite{Gorda:2021kme,Annala:2019puf}, but no reliable results exist in the crucial regime between approximately one and ten nuclear saturation densities. Several model calculations suggest that there is a deconfined phase of quarks. This phase might exist in hybrid stars, i.e., stars made of hadronic matter with an inner core of quark matter. In principle, the transition between the hadron and quark phases can be of first order, of second order or a crossover. At vanishing baryon chemical potential it is quite well established by lattice quantum chromodynamic (QCD) simulations that the transition is a crossover. However, due to the lack of reliable first-principle theoretical methods in the cold and dense regime of the QCD phase diagram, the hadron-quark transition in neutron stars is not fully understood. Indeed, Monte Carlo simulations in lattice QCD are not suitable for $\mu_B/T \gg 1$ due to the fermion sign problem. Nonetheless, progress is being made in that regime with analytically derived effective lattice theories which let open the possibility of a crossover for $\mu_B/T \gg 1$ \cite{Philipsen:2019rjq}. Among other consequences \cite{Baym:2017whm}, the existence of a crossover transition could be a solution of the long-standing hyperon puzzle in neutron star cores \cite{Masuda:2015kha, Masuda:2012ed, Masuda:2012kf}. On the other hand, the possibility of a first-order phase transition cannot be ruled out. 
In fact, some time ago it was conjectured that in the $T$ versus $\mu_B$ plane there may exist a critical point at a critical 
chemical potential $\mu_c$, beyond which the transition is of first order \cite{best}. This critical point has been object of an intense search in the heavy ion collisions at lower energies carried out by the STAR Collaboration in the context of the Beam Energy Scan program \cite{best}. So far, the results are not conclusive. The latest measurements \cite{star23} could be qualitatively  well explained by the ultra-relativistic quantum molecular dynamics (UrQMD) approach, which contains no critical point. However, alternative interpretations were given in  Ref.~\refcite{soren24} and later contested in Ref.~\refcite{olev24}. If the existence of the critical point is confirmed in the future, one may then expect a sharp phase transition in hybrid stars. If not, the phase transition may be smoother.

EOSs with a first order transition can be easily constructed because most descriptions of cold dense matter rely on two different phenomenological models, one for the quark phase and the other for the hadronic phase. Within this kind of analysis the (in general) different functional form of both EOSs induces the phase transition to be of first order. This doesn't mean necessarily that the resulting EOS must be discontinuous. In fact, the outcome depends critically on whether charge neutrality is realized locally or globally in the system (see Ref. \refcite{13b,Hempel:2009vp} and references therein). 
In the first case, the system splits in two homogeneous phases separated by a sharp interface. In the second case, an inhomogeneous system arises where electrically charged lumps of one phase exist in a charged background of the other one \cite{Mariani:2023kdu}. In such a case, the mixed phase EOS connects smoothly a pure low-density hadronic phase with a pure high-density quark phase. In this work, we will make use of an EOS in which the phase transition may be discontinuous (representing a first-order phase transition with local charge neutrality) or smooth (representing either a crossover or a mixed phase).

The EOS of hybrid stars is usually constructed by assuming a configuration consisting of one hadronic matter EOS and one quark matter EOS and then combining them using a Maxwell or a Gibbs construction. 
Depending on the EOSs adopted, the changes in the relevant astrophysical variables may be smoother or sharper during the transition. In this work, we will employ an approach that is unbiased toward any particular model to connect both EOSs. Our aim is to investigate the relationship between the sharpness of the transition and observable quantities.
In the literature there are many works where a similar investigation was conducted \cite{Baym:2017whm,Masuda:2012ed,Masuda:2012kf,Han:2018mtj,Pereira:2022stw}. In Ref.~\refcite{Han:2018mtj} some neutron star observables were calculated with a hybrid EOS which could be tuned to include a sharper or smoother phase transition, depending on the choice of a continuously varying parameter. In Ref.~\refcite{Han:2018mtj} the authors focused on the energy density as a function of the pressure. At the transition point, the energy density jump $\Delta \varepsilon$ and the smoothness were arbitrarily varied. More recently, in Ref.~\refcite{Pereira:2022stw}, a similar analysis was performed. In that work the variable parameter was the pressure gap at the transition, $\Delta P$. 
 
In many works it has been assumed that all the relevant information concerning the phase transition is encoded in the behavior of the speed of sound ($c_S^2$).
In models with a first-order phase transition, $c_S^2$ is initially low on the hadronic side, increases with energy density, drops to zero during the transition, and then rises again in the quark phase, approaching asymptotically the conformal limit of $c_S^2 = 1/3$. Recently, there has been a debate in the literature about the potential occurrence of a ``bump'' in the speed of sound as a function of energy density and whether the conformal limit is approached from above or below. We will address this issue in the present work.

The paper is organized as follows. In Sec. \ref{sec:EOS}, we present the EOSs for pure hadronic matter and pure quark matter, and describe the construction of the crossover or mixed EOS using a Gaussian function to smooth the transition between the two phases. Section \ref{sec:results} details the results of our study, focusing on how the sharpness of the phase transition affects the EOS, the behavior of the speed of sound, and various neutron star properties such as the mass-radius relation and tidal deformability. We also compare our theoretical predictions with existing astrophysical measurements. Finally, in Sec. \ref{sec:conclusions}, we summarize our findings and discuss their implications for the understanding of neutron star structure and the quark-hadron transition.

\section{The equation of state}
\label{sec:EOS}

The EOS for the crossover/mixed phase will be constructed by smoothly matching two EOSs, one for pure hadronic matter and another for pure quark matter, using a Gaussian function. This approach results in a relatively general family of crossover/mixed EOSs, enabling the exploration of a wide range of microscopic possibilities and their potential implications for neutron star structure.

The pure hadronic phase will be characterized by using a generalized piecewise polytropic (GPP) EOS \cite{OBoyle-etal-2020}, which aligns at $1.1\, n_0$ (where $n_0$ is the nuclear saturation density) with predictions based on chiral effective field theory (cEFT) interactions. The cEFT framework provides a systematic expansion of nuclear forces at low momenta, explaining the hierarchy of two-, three-, and weaker higher-body forces \cite{Coraggio:2012ca,Gandolfi:2011xu,Holt:2012yv,Hebeler:2009iv,Sammarruca:2012vb,Tews:2012fj}. Currently, microscopic calculations using cEFT interactions allow for the determination of pressure at $n_0$ with a few percent accuracy.  In the intermediate-density regime above $n_0$, our understanding of the EOS remains limited. However, we will constrain the EOS by imposing causality, thermodynamic consistency, and requiring that the resulting compact stars have a maximum mass above $2 \,M_{\odot}$, in accordance with recent observations of high-mass pulsars.

For the pure quark phase, we will use the vector MIT (vMIT) bag model, which has been widely employed to investigate various neutron star properties. The MIT bag model has successfully reproduced several features of QCD, such as the confinement of quarks inside hadrons and the asymptotic freedom of QCD at high energies. Over the last few years, multiple groups have introduced repulsive interactions among quarks, mediated by the exchange of vector particles \cite{Fogaca:2010mf,Franzon:2012in,Albino:2021zml,Klahn:2015mfa,Cierniak:2018aet,Song:2019qoh,Lopes2021_I, Lopes2021_II, Otto:2020hoz,Pisarski:2021aoz,Su:2021ouy}, which can be ``effective massive gluons'' or ``effective vector mesons''. These mesons significantly influence the quark matter EOS, rendering it sufficiently stiff to support neutron stars with masses above $2 \, M_{\odot}$.

The GPP EOS for pure hadronic matter is described in Section \ref{sec:hadronic_matter}, and the vMIT bag model for pure quark matter in Section \ref{sec:quark_matter}. The mixing/matching between both pure phases is done using a Gaussian prescription, which will be described in detail in Section \ref{sec:crossover}.

\subsection{Pure hadronic matter }
\label{sec:hadronic_matter}

The GPP EOS is implemented as follows. {For the outer regions, we adopt the well-established SLy4 EOS for the hadronic crust, as described in} Ref.~\refcite{Douchin:2001sv}. For the hadronic core portion, we develop two GPP EOSs with differing levels of stiffness. The EOS for this section is divided into three segments by three specific densities. The first boundary, located at $\rho_0$, marks the transition between the crust and the core's lowest-density segment. {The value of $\rho_0 = 10^{14} \text{ g cm}^{-3}$ (corresponding to a baryon number density of approximately $0.059 \text{ fm}^{-3}$) is a standard choice for this transition density.} The densities $\rho_1$ and $\rho_2$ are {not physical transition points but rather the mathematical boundaries between the subsequent polytropic segments of the core. Their values, given in Table \ref{table:hadronic}, were chosen to control the overall stiffness of the EOS while ensuring continuity across the segments.} Between the dividing densities, the rest-mass density $\rho$, energy density $\varepsilon$, and speed of sound $c_S$ are represented by the following functions of pressure $P$:
\begin{eqnarray}
\rho &=& \left(\frac{P - \Lambda_i}{K_i}\right) ^ {\frac{1}{\Gamma_i}}    ,    \\
\varepsilon &=& \frac{K_i \rho ^ \Gamma_i}{\Gamma_i - 1} + (1 + a_i) \rho - \Lambda_i    ,  \\
c_S &= & \left[\frac{1}{\Gamma_i - 1} + \frac{1 + a_i}{K_i \Gamma_i \rho ^ {\Gamma_i - 1}}\right] ^ {-\frac{1}{2}}    ,
\end{eqnarray}
where, following Refs.~\refcite{Hebeler:2013nza}, \refcite{OBoyle-etal-2020}, we use the convention of incorporating the speed of light $c$ into the definition of the energy density and pressure. {In the expressions above, $\Gamma_i$ is the polytropic index and $K_i$ is the polytropic constant for each segment.} As a result, the EOS parameters are given in units such that $\rho$, $\varepsilon$, and $P$ are expressed in $\text{g cm}^{-3}$.
The set of parameters $K_i$, $\Gamma_i$, $a_i$, and $\Lambda_i$ characterize the EOS within each interval $[\rho_{i-1}, \rho_i]$. These parameters are not independent, as continuity and differentiability of the energy density $\varepsilon(\rho)$ and pressure $P(\rho)$ are enforced across the dividing densities (i.e., $c_{S}$ is continuous at these boundaries). 
The parameters $\log_{10} \rho_0$, $\log_{10} K_1$, and $\Gamma_1$, as shown in Table \ref{table:hadronic}, were chosen to ensure that the first segment of the hadronic core EOS {aligns with the pressure and energy density predicted by chiral effective field theory (cEFT) calculations at a density of $1.1 \, n_0$. This serves as a low-density constraint for our phenomenological core EOS, rather than implying the use of a cEFT model directly.}
{While our phenomenological GPP models do not allow for a direct calculation of nuclear saturation properties like the symmetry energy ($E_{sym}$) and its slope ($L$), this limitation is mitigated by the cEFT constraint. The cEFT band is derived from calculations consistent with experimental nuclear properties, so by anchoring our EOSs to it, they effectively inherit this physical consistency. Guided by these constraints, the Soft model was designed to run near the lower boundary of the cEFT band, while the Stiff model follows the upper boundary. Beyond this anchor point, we made the models as soft and as stiff as possible, with the crucial requirement that the purely hadronic EOS must support a maximum mass sufficiently above $2\,M_{\odot}$. This ensures that the final hybrid star, after the softening from the quark phase, still surpasses the observational limit.}

\begin{table}
\centering
\begin{tabular}{c|cccccccc}
\toprule
EOS & $\log_{10}\rho_0$ & $\log_{10}\rho_1$ & $\log_{10}\rho_2$  & $\Gamma_1$ & $\Gamma_2$ & $\Gamma_3$  & $\log_{10}K_1$ \\
\toprule
Soft            & 14 & 14.45 & 14.90 & 3.17 & 3.55 & 3.1  & $-33.210$  \\ 
Stiff           & 14 & 14.74 & 14.99 & 3.18 & 3.52 & 3.1  & $-33.215$  \\
\toprule
\end{tabular}
\caption{Parameters of the Soft and Stiff hadronic EOSs. }
\label{table:hadronic}
\end{table}

\subsection{Pure quark matter}
\label{sec:quark_matter}

In this work, we will use the version of the model presented in Refs.~\refcite{Lopes2021_I}, \refcite{Lopes2021_II}, which is described by the following Lagrangian density:
\begin{equation}
\begin{aligned}
\mathcal{L} & = \sum_{q}\left\{\bar{\psi}_{q}\left[i \gamma^{\mu} \partial_{\mu}-m_{q}\right] \psi_{q}-B\right\} \Theta\left(\bar{\psi}_{q} \psi_{q}\right)  \\
& + \sum_{q} g\left\{\bar{\psi}_{q}\left[\gamma^{\mu} V_{\mu}\right] \psi_{q}\right\} \Theta\left(\bar{\psi}_{q} \psi_{q}\right) +    \tfrac{1}{2} m_{V}^{2} V_{\mu} V^{\mu} \\
& +  \bar{\psi}_{e} \gamma_{\mu}\left(i \partial^{\mu}-m_{e}\right) \psi_{e}   ,
\end{aligned}
\end{equation}
where $q$ denotes quarks ($u$, $d$, and $s$), $e$ denotes electrons, and $B$ is the bag constant that represents the additional energy required to create a perturbative vacuum region per unit volume. The Heaviside step function $\Theta$ equals 1 inside the bag and 0 outside. For simplicity, we adopt here a universal coupling of the quark $q$ with the vector field $V^{\mu}$, with a coupling constant $g$. The mass of the vector field is taken to be $m_{V} = 780 \; \mathrm{MeV}$. 

Working in the mean field approximation and defining $G_{V} \equiv \left({g}/{m_{V}}\right)^{2}$,  one finds the following equation of motion for the vector field:
\begin{equation}
m_{V} V_{0}  = G_V^{1/2} (n_u + n_d + n_s),  
\end{equation}
where $n_{q}=\left\langle\bar{\psi}_{q} \gamma^{0} \psi_{q}\right\rangle$ represents the  quark number density.
Assuming zero temperature, the grand thermodynamic potential per unit volume reads \cite{Lopes2021_II}:
\begin{eqnarray}
\Omega & =&\sum_{q} \Omega^*_{q} + B-\tfrac{1}{2} m_{V}^{2} V_{0}^{2}  + \Omega_e
\label{eq:Omega_bulk}
\end{eqnarray}
being $\Omega^*_{q}$ the grand thermodynamic potential of a free Fermi gas of quarks 
with the following effective chemical potential:
\begin{eqnarray}
\mu_{q}^{*} &=&  \mu_{q} -  G_V^{1/2} m_V V_{0} .
\label{eq:effective_mu}
\end{eqnarray}
The EOS can be easily derived from Eq.~\eqref{eq:Omega_bulk} and its explicit form is presented in Refs.~\refcite{Lopes2021_I}, \refcite{Lopes2021_II}. To apply these equations to neutron stars, we assume that quarks and electrons are in chemical equilibrium under weak interactions and impose charge neutrality and baryon number conservation. We adopt $m_{u} = m_d = 4$ MeV, $m_{s}=$ $95$ MeV, and electrons are assumed to be massless. {The parameters $B$ and $G_V$ are treated as free parameters. The bag constant $B$ represents the energy cost to deconfine hadrons, acting as a negative pressure term. A larger $B$ delays the phase transition to higher densities. The vector coupling $G_V$ introduces repulsion between quarks, making the EOS stiffer and able to support more massive stars. The specific values in Table \ref{table:hybrid} were chosen as representative examples that ensure the final hybrid EOS is consistent with observations of pulsars with masses $\gtrsim 2\,M_{\odot}$. While different choices would alter the quantitative results, our qualitative conclusions regarding the effect of the transition's sharpness are expected to hold.}

\subsection{Crossover/mixed equation of state}
\label{sec:crossover}

\begin{figure}[tb]
\centering 
\includegraphics[width=0.55\columnwidth]{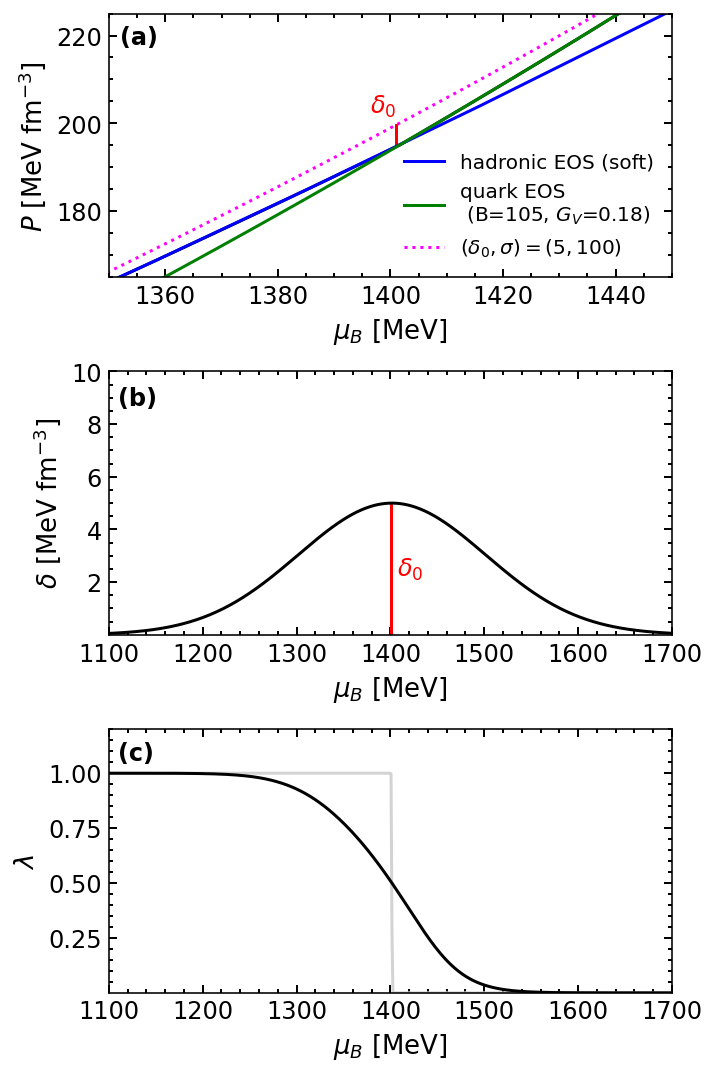}
\caption{Physical interpretation of $\delta(\mu_B)$ and $\lambda(\mu_B)$. The curves were obtained with the Soft hadronic EOS and the vMIT EOS with $B=105 ~ \text{MeV fm}^{-3}$ and $G_V = 0.18 ~ \mathrm{fm^2}$. Both EOSs intersect at $\mu_c = 1401 ~\mathrm{MeV}$ in the $P$ versus $\mu_B$ plane. The dotted line represents the crossover/mixed EOS
with parameters $\delta_0 = 5 ~ \text{MeV fm}^{-3}$ and $\sigma = 100 ~  \mathrm{MeV}$. The maximum of $\delta(\mu_B)$ occurs at $\mu_c$.}
\label{fig:delta}
\end{figure}

To derive hybrid EOSs featuring phase transitions with varying degrees of sharpness, we combine hadronic and quark matter EOSs using a Gaussian function. The total pressure $P(\mu_{B})$ is determined from the pressure of the pure phases ($P_{H}(\mu_{B})$ and $P_{Q}(\mu_{B})$) through the following relationship\footnote{This formula is inspired by an expression presented in Ref.~\refcite{Hama:2005dz} in the context of heavy-ion collisions.}:
\begin{align}
\left(P-P_{Q}\right)\left(P-P_{H}\right)=\delta^2(\mu_{B}).
\label{eq:start}
\end{align}
The Gaussian function $\delta$ is centered at a chemical potential $\mu_c$ and has a width $\sigma$: 
\begin{align}
\delta(\mu_B) & = \delta_0 \exp\left[ - \frac{(\mu_B - \mu_{c})^2}{2 \sigma^2} \right].
\label{delb}
\end{align}
One can choose the value of $\mu_c$ symmetrically by setting it equal to  the chemical potential where the pressures $P_{H}(\mu_{B})$ and $P_{Q}(\mu_{B})$ of the pure phases intersect. Alternatively, one can choose it asymmetrically by selecting a value of $\mu_c$ that is either larger or smaller than the intersection chemical potential. The width of the region where the crossover transition takes place is determined by the value of $\sigma$. The prefactor $\delta_0 = \delta(\mu_c)$ quantifies how much the crossover EOS deviates from the pure phases at $\mu_B = \mu_c$. When $\mu_c$ is chosen symmetrically, such that $P_{H}(\mu_{c}) = P_{Q}(\mu_{c})$, then $\delta_0$ represents the overpressure of the crossover EOS compared to the pure phases at their intersection (see Fig.~\ref{fig:delta}a).

It is crucial to note that the parameters $\delta_0$ and $\sigma$ are interrelated and cannot be selected arbitrarily. This can be understood from Fig.~\ref{fig:delta}a. If we set $\sigma$ and increase $\delta_0$ to a sufficiently large value, the crossover curve will become concave downward, that is, $\partial^2 P/ \partial \mu^2 < 0$. This situation is not physically acceptable since it implies that the speed of sound, $c_S$, given by
\begin{align}
\frac{1}{c_S^2}=\frac{\mu}{n} \frac{\partial^2 P}{\partial \mu^2},
\end{align}
would be negative. To prevent this unphysical behavior, the width $\sigma$ of the crossover region must be large enough for a given $\delta_0$. Similarly, if the width $\sigma$ is small, the value of $\delta_0$ must be sufficiently small (see Fig.~\ref{fig:delta}b). 

{Based on this interplay, the specific pairs of ($\delta_0, \sigma$) used in this work were chosen to represent a spectrum of transition sharpnesses, from nearly first-order to a very smooth crossover. The chosen sets---namely $(\delta_0 [\text{MeV fm}^{-3}], \sigma [\text{MeV}]) = (2, 50)$, $(5, 100)$, and $(10, 150)$---all respect the causality constraint while allowing us to study the impact of progressively smoother transitions on the EOS and stellar properties.}

From Eq.~\eqref{eq:start}, we can express $P$ in the following way:
\begin{align}
P = \lambda P_H +(1 -\lambda) P_Q  + \frac{2 \delta^2 }{\sqrt{(P_Q -P_H)^2 +4\delta^2 }},  
\label{eq:pressure}
\end{align}
being
\begin{align}
\lambda(\mu_B) = \tfrac{1}{2} \left[1 -\frac{P_Q -P_H}{\sqrt{(P_Q -P_H)^2 +4\delta^2}} \right].
\label{eq:lambda} 
\end{align}
The quantity $\lambda$ exhibits a Heaviside step function-like behavior around $\mu_B = \mu_c$, as shown in Fig.~\ref{fig:delta}c. When $\mu_B \ll \mu_c$, one finds $\lambda \approx 1$ and $\delta^2 \approx 0$, indicating that the system is composed of pure hadronic matter. When $\mu_B \gg \mu_c$, then $\lambda \approx 0$ and $\delta^2 \approx 0$, indicating that the system is composed of pure quark matter. For $\mu_B \sim \mu_c$, one obtains $0 < \lambda < 1$ and $\delta^2 > 0$, indicating that the system is a combination of hadronic and quark matter.

The baryon number density $n_B \equiv \partial P /\partial \mu_B$ is given by:
\begin{align}
n_B = \lambda n_{B,H} +(1 -\lambda) n_{B,Q}  + \frac{(\delta^2)^\prime }{\sqrt{(P_Q -P_H)^2 +4\delta^2}},
\label{eq:nB}
\end{align}
being
\begin{align}
(\delta^2)^\prime   =  -\frac{(\mu_B - \mu_{c})}{2 \sigma^2} \delta^2. 
\label{eq:delta_prime} 
\end{align}
The energy density is obtained from Euler's relation $\varepsilon = -P + \mu_B n_B$.

\begin{table}[tb]
\centering
\begin{tabular}{c|c|cc}
\toprule
Hybrid EOS & Hadron EOS & \multicolumn{2}{c}{Quark EOS} \\
\cline{3-4} 
           &            & $B$ ($\text{MeV fm}^{-3}$) & $G_V$ (fm$^{2}$) \\
\toprule
Model 1    & Soft       & 105              & 0.18 \\
Model 2    & Stiff      & 100              & 0.21 \\
\toprule
\end{tabular}
\caption{Hybrid EOS models used in the calculations, with corresponding parameters for the hadronic and quark phases. Parameters for the hadronic EOS (`Soft' and `Stiff') are detailed in Table \ref{table:hadronic}. The quark EOS is fully determined by speci\-fying the bag constant $B$ and the vector coupling constant $G_V$.}
\label{table:hybrid}
\end{table}

\section{Results}
\label{sec:results}

\begin{figure}[tb]
\centering 
\includegraphics[width=0.65\columnwidth]{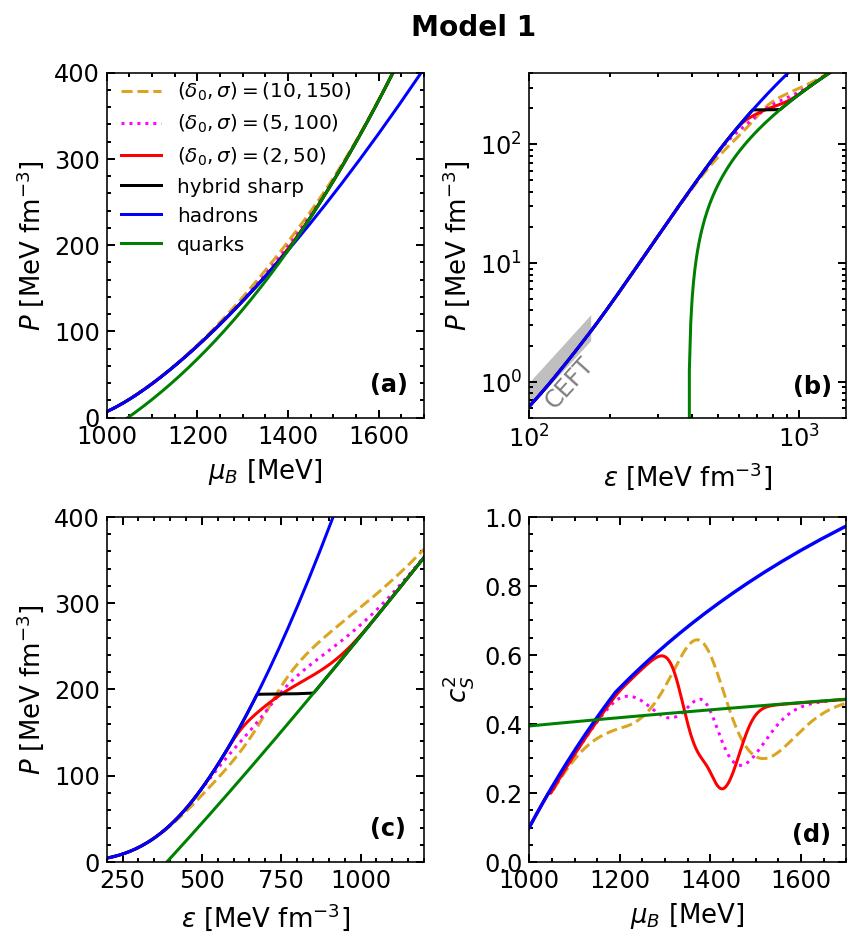}
\caption{The results were obtained by using the Soft hadronic EOS and the vMIT EOS with $B=105 ~ \text{MeV fm}^{-3}$ and $G_V = 0.18 ~ \mathrm{fm^2}$. In all cases we used $\mu_c = 1401 ~\mathrm{MeV}$, which is the baryon chemical potential at which the pressure of the pure phases intersect. For the parameters ($\delta_0 [\text{MeV fm}^{-3}]$, $\sigma [ \mathrm{MeV}]$) we used: (2, 50), (5, 100) and (10,150). We also show the sharp (first-order) phase transition. }
\label{fig2}
\end{figure}

\begin{figure}[tb]
\centering 
\includegraphics[width=0.65\columnwidth]{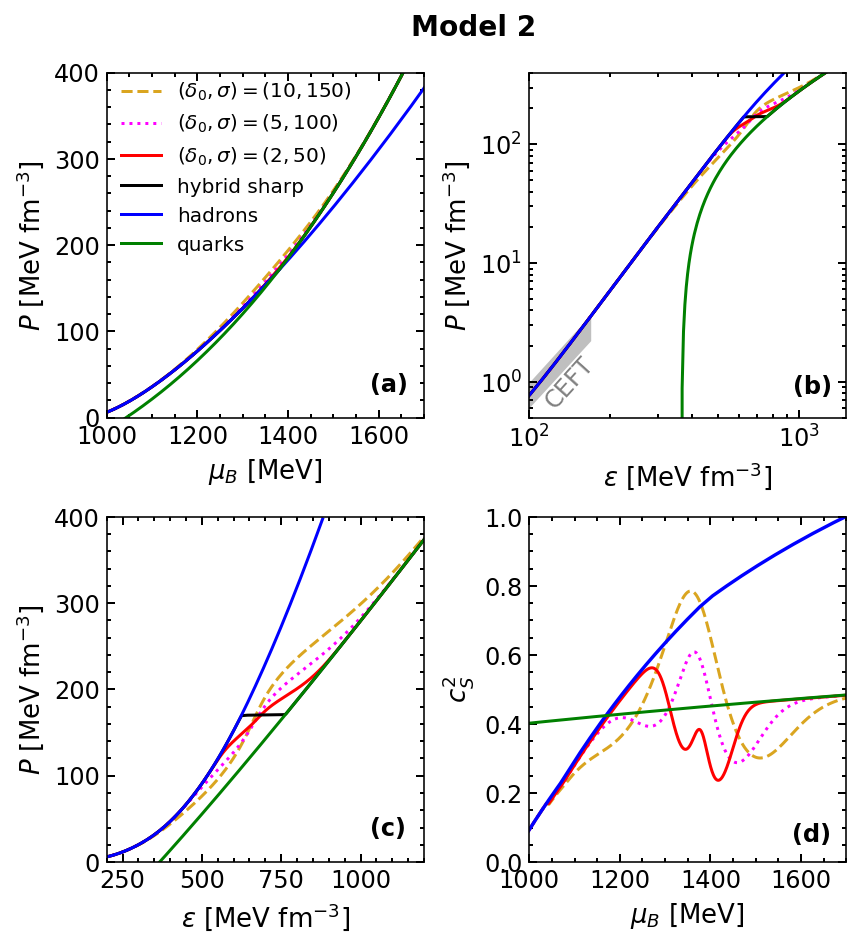}
\caption{Same as in the previous figure but using the Stiff hadronic EOS and the vMIT EOS with $B=100 ~ \text{MeV fm}^{-3}$ and $G_V = 0.21 ~ \mathrm{fm^2}$. In this case we obtain $\mu_c = 1378 ~\mathrm{MeV}$.}
\label{fig3}
\end{figure}

In Figs.~\ref{fig2} and \ref{fig3}, we present an analysis of the hybrid EOSs constructed using the ansatz detailed in the previous section. To explore phase transitions of varying sharpness, we selected different values for $\delta_0$ and $\sigma$. The parameter pairs employed  are $(\delta_0 [\text{MeV fm}^{-3}], \sigma [ \mathrm{MeV}]) = (2, 50)$, $(5, 100)$, and $(10, 150)$.
In Fig.~\ref{fig2}, we use a quark EOS parametrization with $B=105 ~ \text{MeV fm}^{-3}$ and $G_V = 0.18 ~ \mathrm{fm^2}$ next to a soft hadronic EOS. Meanwhile, Fig.~\ref{fig3} employs an alternative quark EOS parametrization with $B=100 ~ \text{MeV fm}^{-3}$ and $G_V = 0.21 ~ \mathrm{fm^2}$ alongside a stiff hadronic EOS (see Table \ref{table:hybrid}). 
{Each figure is divided into four panels, each highlighting a different aspect of the hybrid EOS. Panel (a) in each figure displays the pressure versus the baryon chemical potential. Panel (b) illustrates the pressure as a function of energy density ($P$ vs. $\varepsilon$) on a logarithmic scale to show the global behavior, including the cEFT constraints. Panel (c) also shows $P$ vs. $\varepsilon$, but on a linear scale and zoomed into the phase transition region to better visualize its details. Finally, panel (d) shows the speed of sound as a function of the baryon chemical potential.}

{At first glance, the main visual difference between the sets of curves in Figs.~\ref{fig2} and \ref{fig3} appears in the speed of sound (panel d). However, differences are also visible in panel (c), stemming mainly from the combination of distinct EOS for quarks and hadrons in each model, which results in a different phase transition pressure and energy density jump. Since the same smoothing procedure is applied to these different underlying transitions, the qualitative effect of the smoothing parameters ($\delta_0, \sigma$) appears similar in both figures. Despite this, the underlying quantitative differences are appreciable, and they are more clearly visible in the P-$\varepsilon$ plot (panel c), which uses a linear scale, than in panels (a) and (b). These quantitative differences in the EOS, lead to significant variations in its derivative ($\partial P / \partial \varepsilon$), which are dramatically visualized in the speed of sound plots.}

A sharp transition is achieved by setting $\delta \to 0$. As $\delta$ increases, the phase transition gradually becomes smoother, progressively eliminating the plateau in the $P$ versus $\varepsilon$ plot.
An important point to note is that the pressure difference between the sharp transition model and the models with smooth transitions considered here lies within the range of $2-10 \, \text{MeV fm}^{-3}$ when $\mu_B$ reaches its critical value $\mu_c$. These values might appear relatively small when compared to the pressure $P(\mu_c)$, which is approximately $200 \, \text{MeV fm}^{-3}$. Notably, these pressure differences cannot be significantly increased because, as discussed in the previous section, the width of the mixed phase or crossover region would have to increase if the pressure difference $\delta_0$ were to be raised. Consequently, upon inspecting Figs.~\ref{fig2}a and \ref{fig3}a, it might seem that the smoothing of the EOS should not lead to significant changes in the astrophysical context.   

However, this is not what actually occurs. First, the most important factor in determining the stellar structure is the relationship between pressure and energy density. As shown in Figs.~\ref{fig2}b,c and \ref{fig3}b,c, the pressure differences between the different parametrizations can reach some tens of $\text{MeV fm}^{-3}$ for certain values of $\varepsilon$. This has a significant effect on the stiffness of the EOS, which greatly impacts the mass-radius relationship and other global properties of neutron stars. Similar to what occurs in microscopic models of the quark-hadron pasta phase \cite{Mariani:2023kdu}, the EOS becomes softer for pressures below the transition pressure and stiffer for $P > P(\mu_{c})$.  

Secondly, the considerable width of the hybrid region also plays a critical role in shaping the EOS. For the case $\delta_0 = 10 \, \text{MeV fm}^{-3}$, the width of the hybrid region is substantial, extending from 500 to $1500 \, \text{MeV fm}^{-3}$, which represents a vast range of stellar core densities. The widening of the hybrid region observed here exhibits similar characteristics to those found in models of the quark-hadron pasta phase (see, for example, Fig. 1 of \cite{Mariani:2023kdu}). Much higher values of $\delta_0$ do not seem reasonable, as they would imply that the mixed phase/crossover would begin to intrude in the region near the nuclear saturation density, where no form of quark matter is expected.

\begin{figure*}[tb]
\centering 
\includegraphics[width=0.65\columnwidth]{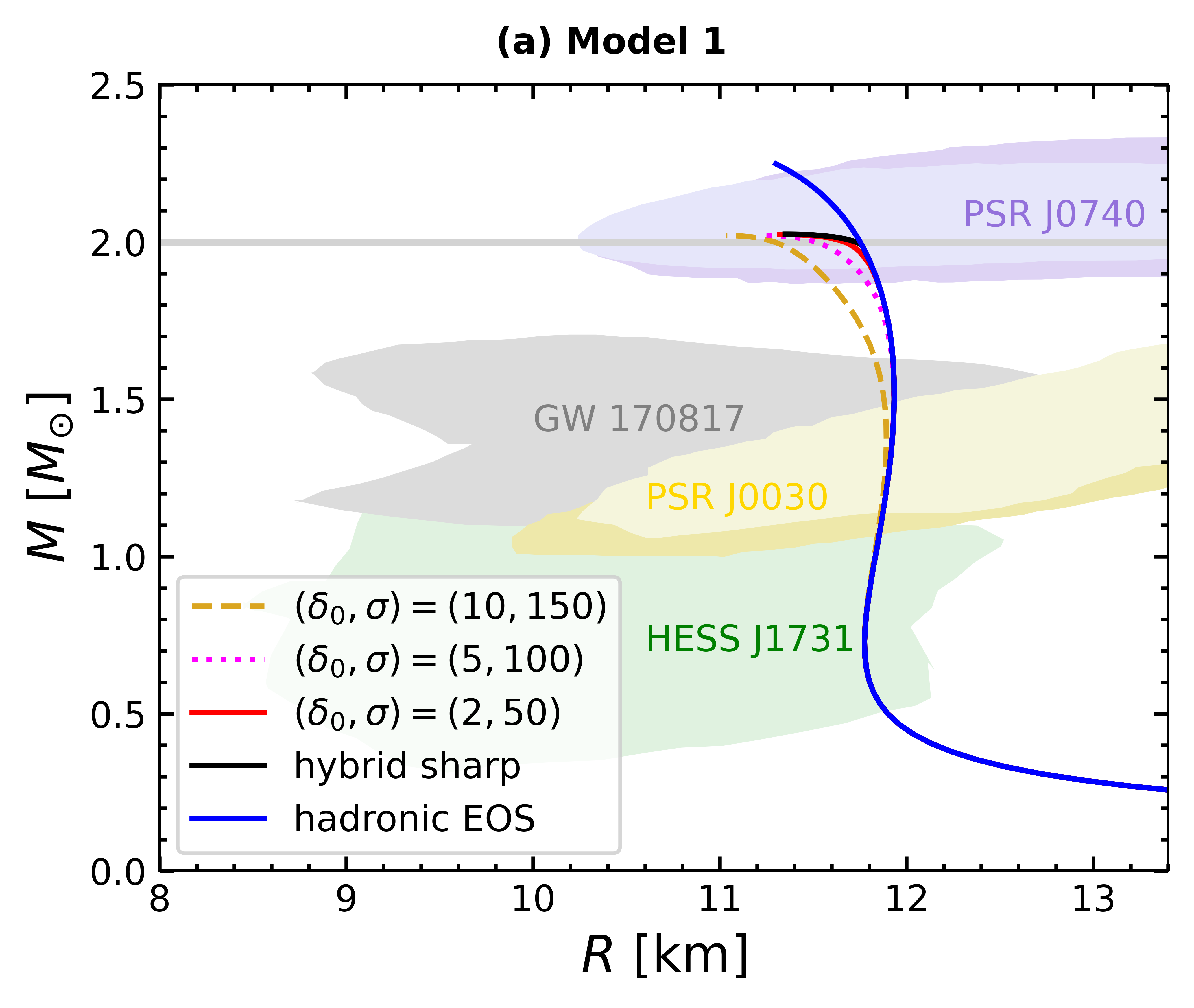}
\includegraphics[width=0.65\columnwidth]{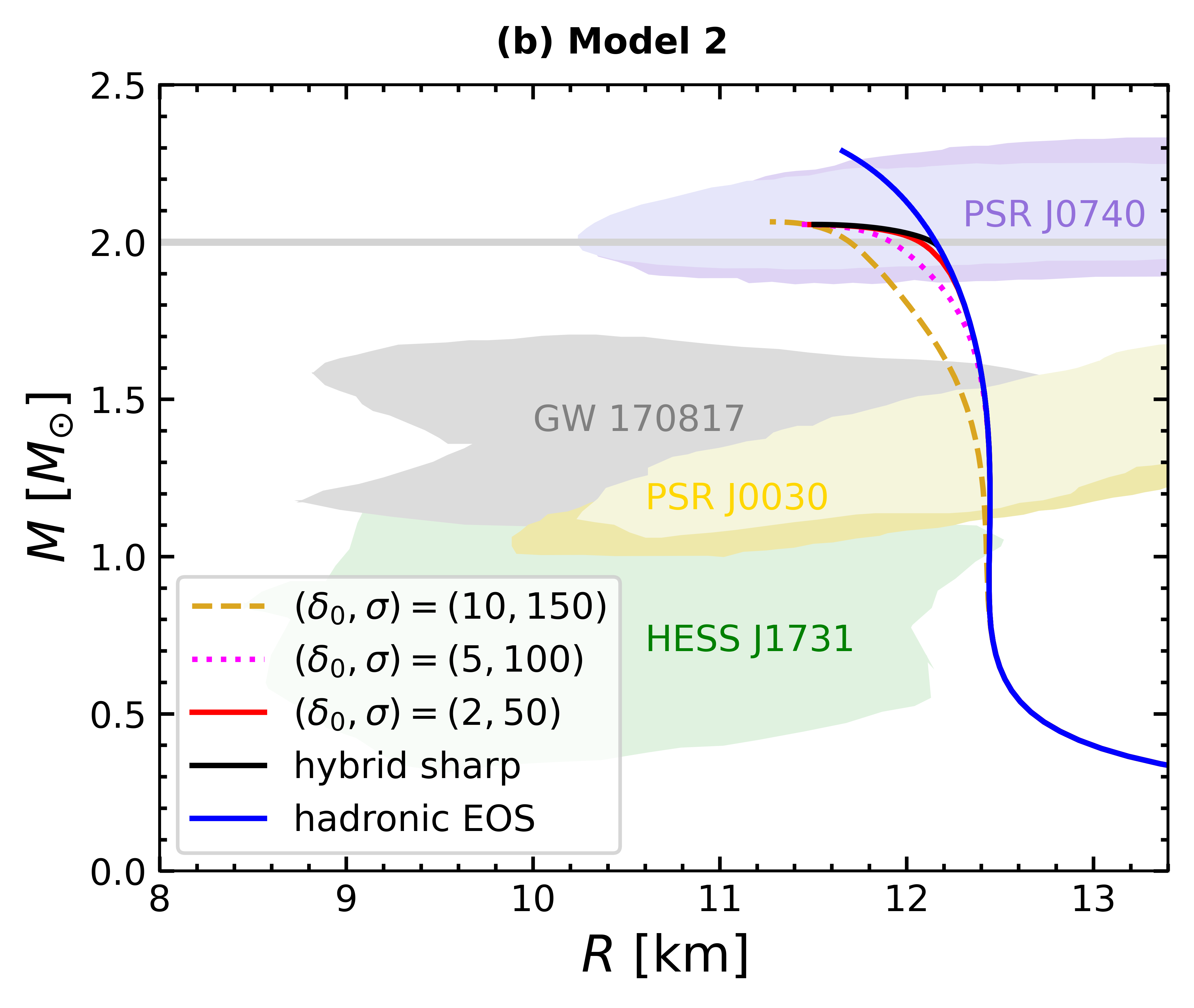}
\caption{Mass-radius relation for the EOSs shown in Figs.~\ref{fig2} and \ref{fig3}: (a) Model 1 and (b) Model 2. Current astrophysical constraints are also included for reference (details provided in the text).}
\label{fig:tov}
\end{figure*}

An interesting aspect of the model discussed here is the wide variety of behaviors exhibited by the speed of sound. 
{This complex behavior, featuring prominent peaks and valleys, is a direct consequence of our interpolation formalism.  Our formula for the pressure (Eq.~\eqref{eq:pressure}) involves a non-linear mixing of the hadronic and quark phases, and the derivatives of this expression lead to complex terms that naturally result in non-monotonic behavior. It is important to emphasize that these peaks do not represent the appearance of new particles or degrees of freedom. Rather, they are features of this phenomenological model that reflect a rapid change in the stiffness of matter as it transitions from being hadron-like to quark-like. A peak in $c_S^2$ simply indicates a region where the matter is locally very stiff.} 
In the model with the parametrization (2, 50), the speed of sound experiences a significant drop near the first-order phase transition region (Figs.~\ref{fig2}d and \ref{fig3}d). During this drop, the curve does not gradually approach the speed of sound of quark matter. Instead, its value continues to decrease significantly, reaching approximately half the speed of sound in quark matter. From there, it begins to increase gradually, approaching from below the quark speed of sound value, which is approximately $0.4c$ (Fig.~\ref{fig2}d). In the case of Fig.~\ref{fig3}d, there is also a secondary peak following the abrupt drop, adding further complexity to the behavior of the speed of sound.
The model with the parametrization (10, 150) also exhibits an interesting and complex behavior in $c_S$. At low densities, the curve corresponding to this model closely follows the hadron matter curve, but then gradually and smoothly diverges from it. However, in the intermediate regime, near the first-order phase transition region, the curve rises significantly and can reach values of $c_S$ that are even higher than those of the hadronic model, forming a well-defined peak. After this peak, $c_S$ shows a qualitatively similar behavior to that of the model (2, 50), displaying a significant drop followed by a valley and a gradual convergence to the quark matter speed of sound.
The model with the parametrization (5, 100) also exhibits distinctive behavior in the intermediate region. Near the first-order phase transition, unlike the abrupt drop observed in the model (2, 50) and the pronounced peak in the model (10, 150), the model (5, 100) shows a greater number of peaks and valleys, though they are less prominent. After this oscillatory regime, the curve gradually converges towards the speed of sound of quark matter, consistently approaching from below.

Note that due to the repulsive vector interactions, $c_S^2$ does not asymptotically approach the conformal limit of $1/3$, but instead converges to a value above it. This is not necessarily a problem, as the model is being applied at densities much lower than those where asymptotic freedom is expected. In a more sophisticated formulation, it would be possible to ensure that the speed of sound asymptotically approaches the conformal value. For instance, this could be achieved by making the coupling $G_V$ density-dependent, allowing it to vanish at very high densities. This approach was recently implemented in Ref.~\refcite{Pinto:2022elo} within the framework of the Nambu-Jona-Lasinio model.

\begin{figure*}[tb]
\centering 
\includegraphics[width=0.65\columnwidth]{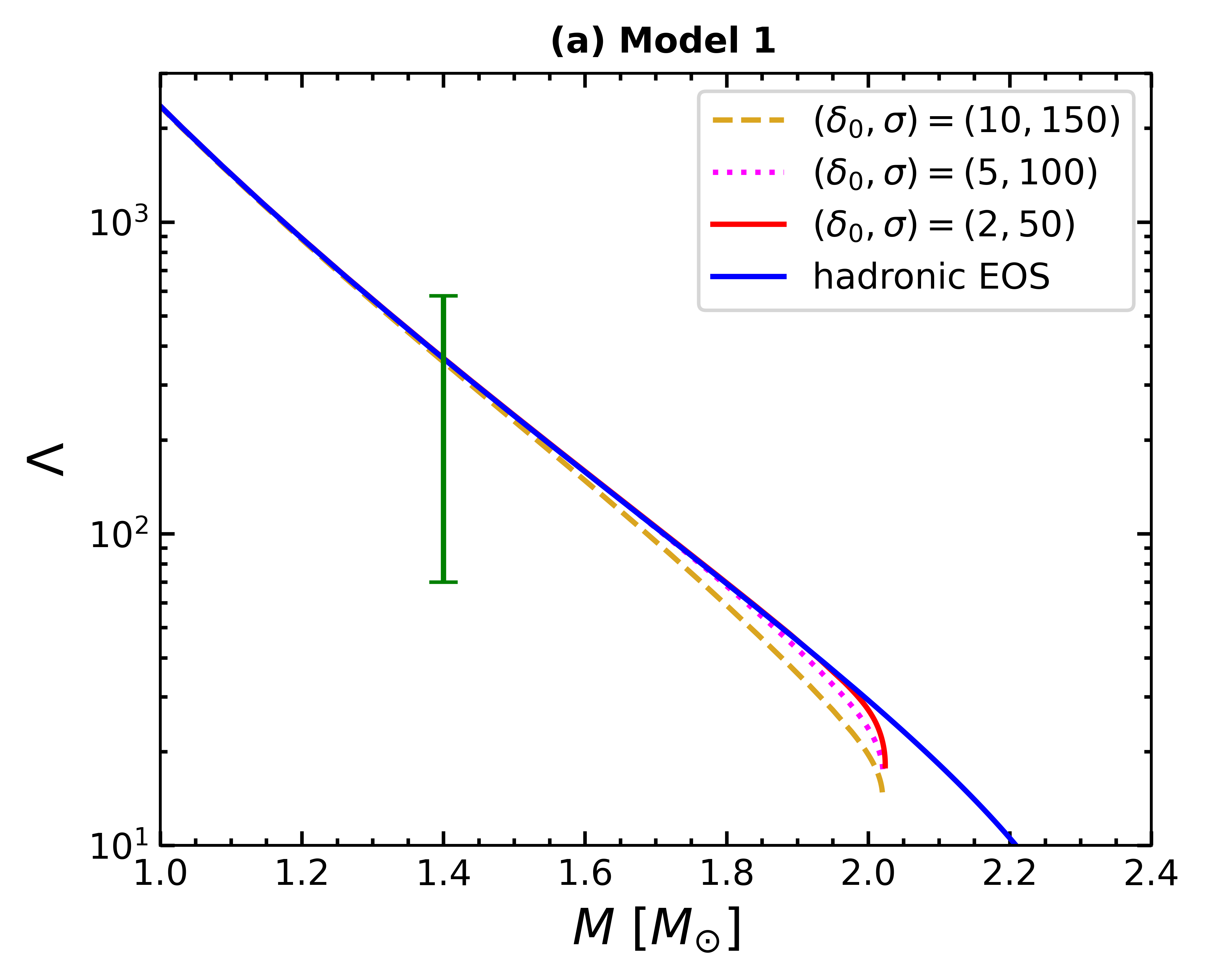}
\includegraphics[width=0.65\columnwidth]{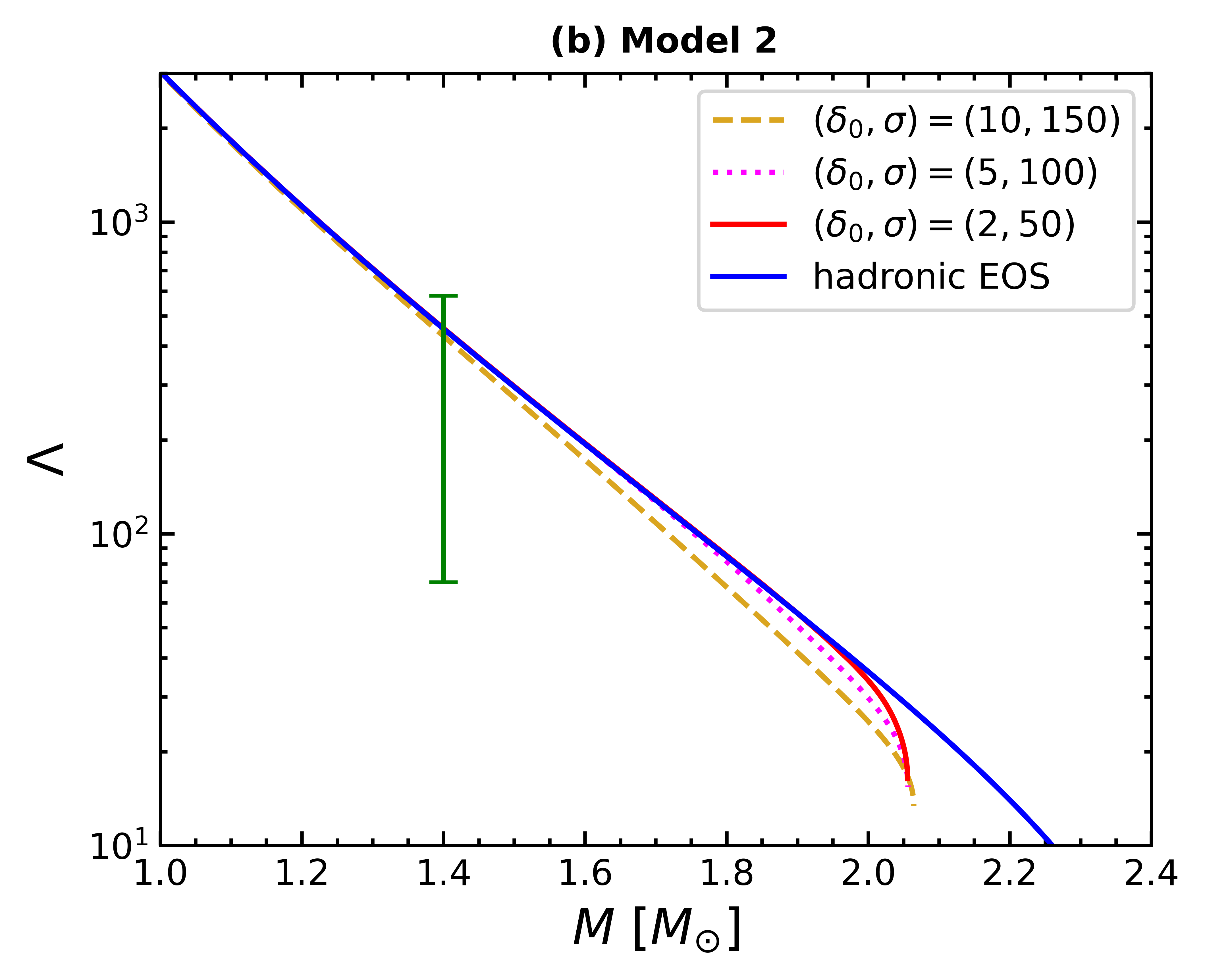}
\caption{Dimensionless tidal deformability as a function of stellar mass for the EOSs shown in Figs.~\ref{fig2} and \ref{fig3}: (a) Model 1 and (b) Model 2.  The vertical bar represents observational data from the GW170817 event detected by LIGO/Virgo, as referenced in the text.}
\label{fig:tidal}
\end{figure*}

To describe the structure of static (non-rotating) compact stars, we solved the Tolman-Oppenheimer-Volkoff equations alongside the equations for tidal deformability \cite{Hinderer:2007mb, Damour:2009vw, Postnikov:2010yn}.     

Figure~\ref{fig:tov} presents our mass-radius relationship calculations alongside current astrophysical constraints, including those from the $\sim 2M_{\odot}$ pulsars \cite{Demorest:2010bx, Antoniadis:2013pzd, Cromartie:2019kug, Linares:2018ppq}, the GW170817 event at the 90\% posterior credible level \cite{LIGOScientific:2018hze, LIGOScientific:2018cki}, NICER pulsars at the 2$\sigma$ posterior credible level \cite{Riley:2019yda, Miller:2019cac, Riley:2021pdl, Miller:2021qha}, and HESS J1731-347 at the 2$\sigma$ posterior credible level \cite{HESS2022}.

The results indicate that variations in the sharpness of the phase transition have a minimal impact on the maximum mass of the star. As shown in Fig. \ref{fig:tov}, the maximum mass remains almost unchanged across models with a sharp transition and those with parametrizations (2,50), (5,100), and (10,150). However, the sharpness of the phase transition plays a significant role in determining the stellar radius.  As the density range of the hybrid region broadens, we observe essentially two effects. First, the part of the curve representing stars containing hybrid matter appears earlier, i.e., at lower mass values. This can be observed in the curve with parametrization (10,150), where the hybrid branch emerges at $\sim 1 M_{\odot}$. In contrast, for less smooth transitions, such as the (2,50) parametrization, hybrid stars are only observed for masses greater than $\sim 1.8 M_{\odot}$. Notably, this latter parametrization exhibits behavior closely resembling that of the sharp transition (cf. Fig. \ref{fig:tov}).   

Secondly, the smoother and wider the density range covered by the hybrid phase, the more the curve shifts to the left in the mass-radius diagram. For example, in the model with parametrization (10,150), the star with a mass of $\sim 1.8 M_{\odot}$ has a radius approximately half a kilometer smaller than that of a purely hadronic star of the same gravitational mass. In a more extreme case, for the same parametrization (10,150), the star with maximum mass has a radius approximately 1 km smaller than the hadronic star of the same gravitational mass. Conversely, less smooth parametrizations with narrower density ranges lead to significantly smaller differences in radii.

{It is important to address the large difference in radii between the stellar sequences shown in Fig.~\ref{fig:tov}(a) (Model 1) and Fig.~\ref{fig:tov}(b) (Model 2). This difference is not an effect of the transition sharpness itself, but rather a direct consequence of the different underlying hadronic EOSs used: Soft for Model 1 and Stiff for Model 2. A stiffer EOS, by definition, provides more pressure support at a given density, resulting in larger stellar radii. Our goal in presenting both models was precisely to demonstrate that the qualitative effect of smoothing the phase transition—namely, the leftward shift of the hybrid branch on the $M-R$ diagram—is a general feature that occurs irrespective of the initial stiffness of the hadronic matter EOS.}

As indicated by the shaded areas in Fig.~\ref{fig:tov}, current observational capabilities are insufficient for distinguishing variations in radius of this magnitude. However, significant improvements in the resolution of radius measurements are anticipated in the coming years, particularly from observatories like NICER. It is important to highlight, though, that if the phase transition’s sharpness is not very pronounced, detecting this feature directly may become challenging, as demonstrated by the minimal impact of the (2,50) parametrization on the mass-radius relationship compared to the traditional first-order (Maxwell construction) scenario.

In Fig. \ref{fig:tidal}, we show the tidal deformation for the same models presented in the previous figures, along with the observational constraint from the GW170817 event \cite{LIGOScientific:2018hze, LIGOScientific:2018cki}. The qualitative shape of the curves with a smooth transition is the same as that of any curve for hadronic matter, that is, \(\Lambda\) is a decreasing function of the stellar mass. The effect of smoothing the phase transition is visible in the fact that \(\Lambda\) is slightly reduced compared to the hadronic model and the models with a sharp transition. This effect is more noticeable for higher-mass objects and becomes almost imperceptible in the region around \(1.4 M_{\odot}\), which is where the GW170817 event could potentially  constrain the theoretical predictions.

\section{Summary and Conclusions}
\label{sec:conclusions}

In this work, we investigated the effects of the sharpness of the phase transition on various aspects of neutron stars. To construct a hybrid EOS with different levels of sharpness, we combined two distinct EOSs: one for hadronic matter and another for quark matter. These were mixed using a Gaussian function, ensuring that the matter is purely hadronic at low densities, purely quark at high densities, and a mixture of both at intermediate densities. The blending of the two EOSs is symmetric around the first-order phase transition point. The Gaussian mixing introduces a parameter \(\delta_0\), which represents the overpressure of the hybrid EOS compared to the pressure of the pure phases at the baryon chemical potential corresponding to the first-order phase transition.
It is important to emphasize that the parameter $\sigma$ (which relates to the range of $\mu_B$ over which the hybrid region extends) and the parameter $\delta_0$ are interconnected and cannot be chosen independently. Setting $\sigma$ and increasing $\delta_0$ arbitrarily is not feasible because, beyond a certain point, this would cause the speed of sound to become negative. Practically, this constraint limits the values of $\delta_0$ from being significantly greater than $10 \, \text{MeV fm}^{-3}$. For typical EOSs, exceeding this limit would result in an excessively wide hybrid region that would start to encroach upon the region around nuclear saturation density, where the presence of quark matter is not expected.

Regarding the relationship between pressure and energy density, we found the following: First, similar to what occurs in microscopic models of the quark-hadron pasta phase \cite{Mariani:2023kdu}, the EOS becomes softer for $P < P(\mu_{c})$ and stiffer for $P > P(\mu_{c})$. Additionally, we observed significant variations in stiffness among the different parametrizations of the model. For instance, in the case of the smoothest parametrization analyzed in this work, $(\delta_0 [\text{MeV fm}^{-3}], \sigma [\text{MeV}]) = (10, 150)$, the pressure can be several tens of $\text{MeV fm}^{-3}$ above or below what is predicted by the sharp first-order phase transition EOS. Furthermore, the aforementioned smooth transition spans a density range very similar to that found in analyses based on microscopic models \cite{Mariani:2023kdu}.   

The speed of sound exhibits a wide variety of behaviors. In some parametrizations of the sharpness, the speed of sound curve shows drastic drops, in other cases, it has pronounced peaks, and in yet others, it displays a more oscillatory pattern with multiple peaks and valleys. This diversity in the speed of sound behavior is significant in the context of studying non-radial oscillations of neutron stars, as different functional forms of $c_S$ can lead to markedly different behaviors in the frequencies and damping rates of oscillation modes, which have direct implications for the gravitational wave emissions from oscillating neutron stars.

In the context of stellar structure, we found that variations in the sharpness of the phase transition have minimal impact on the maximum mass in the mass-radius relationship but can significantly affect the stellar radii. As the phase transition becomes smoother and the density range of the hybrid region broadens, the portion of the curve representing stars with hybrid matter appears at progressively lower mass values and shifts to the left on the $M-R$ diagram. In some cases, this lateral shift of the $M-R$ curve can be substantial (up to 1 km); however, the resolution of current observational capabilities is insufficient to detect radius variations of this magnitude.   {It is also worth noting how the presence of other degrees of freedom, such as hyperons, would affect our results. The inclusion of hyperons is known to soften the hadronic EOS, which would generally lower the density and pressure at which the transition to quark matter occurs. Our formalism for smoothing the transition could be applied directly to such a hyperonic EOS. We would expect the same qualitative behavior discussed in this work: a smoother transition would lead to smaller radii for the resulting hybrid hyperonic stars compared to a sharp transition, although the entire mass-radius sequence would likely be shifted towards lower maximum masses due to the initial softening of the hadronic phase.}  
We also analyzed the effect of the sharpness of the phase transition on tidal deformability. The tidal deformability is a decreasing function of stellar mass, exhibiting the same behavior known in hadronic stars and those with a sharp phase transition. As the phase transition becomes smoother, we observe a slight reduction in $\Lambda$, especially for higher-mass objects, compared to the hadronic and sharp transition models. However, this shift is significantly smaller than the error bar of the GW170817 event.

The effects described above are quite general {in a qualitative sense. This means that the main trends do not depend on the stiffness of the specific EOSs used: a smoother transition consistently makes the EOS softer for pressures below the transition point and stiffer for pressures above it, leading to a leftward shift of the hybrid star branch in the mass-radius diagram.} In fact, the mass-radius relations obtained in state-of-the-art microscopic models of stars with quark-hadron pasta phases \cite{Mariani:2023kdu} (which include the self-consistent calculation of surface and curvature tensions \cite{Lugones:2021tee, Lugones:2016ytl, Lugones:2018qgu}) exhibit exactly the same {qualitative} behavior as those described above.
Other parametric methods for smoothing the phase transition tend to yield similar results in the key aspects \cite{Han:2018mtj, Pereira:2022stw, Cierniak:2021knt}.

\section*{Acknowledgments}
R.~F. receives funding from CNPq (grant 173012/2023-0).\ G.~L. acknowledges the partial financial support from CNPq (grant 316844/2021-7) and FAPESP (grant 2022/02341-9).

\bibliographystyle{ws-ijmpd}
\bibliography{references}

\end{document}